\documentclass[twocolumn]{aastex63}
\usepackage{natbib}
\usepackage{color}
\usepackage{mathptmx}
\usepackage{amsmath}
\usepackage{url}
\usepackage{multirow}
\usepackage{verbatim}

\usepackage[T1]{fontenc}
\usepackage[flushleft]{threeparttable}
\usepackage{hyperref}

\graphicspath{{./}{figures/}}

\shorttitle{{\it AGORA} Comparison. Public data release 2}
\shortauthors{{\it AGORA} Collaboration et al.}

\begin{document}

\title{THE {\it AGORA} HIGH-RESOLUTION GALAXY SIMULATIONS COMPARISON PROJECT:  COSMORUN PUBLIC DATA RELEASE}

\author[0000-0002-6299-152X]{Santi Roca-F\`{a}brega}
\affil{Lund Observatory, Division of Astrophysics, Department of Physics, Lund University, SE-221 00 Lund, Sweden}
\affil{Departamento de F\'{i}sica de la Tierra y Astrof\'{i}sica, Facultad de Ciencias F\'{i}sicas, Plaza Ciencias, 1, 28040 Madrid, Spain; \rm{\href{mailto:santi.roca_fabrega@fysik.lu.se}{santi.roca\_fabrega@fysik.lu.se}}}

\author[0000-0003-4464-1160]{Ji-hoon Kim}
\affiliation{Institute for Data Innovation in Science, Seoul National University, Seoul 08826, Korea; \rm{\href{mailto:mornkr@snu.ac.kr}{mornkr@snu.ac.kr}}}
\affiliation{Center for Theoretical Physics, Department of Physics and Astronomy, Seoul National University, Seoul 08826, Korea}
\affiliation{Seoul National University Astronomy Research Center, Seoul 08826, Korea}

\author[0000-0001-5091-5098]{Joel R. Primack}
\affil{Department of Physics, University of California at Santa Cruz, Santa Cruz, CA 95064, USA; \rm{\href{mailto:joel@ucsc.edu}{joel@ucsc.edu}}}

\author{Anna Genina}
\affil{Max-Planck-Institut f\"{u}r Astrophysik, Karl-Schwarzschild-Str. 1, D-85748, Garching, Germany}

\author[0000-0002-9144-1383]{Minyong Jung}
\affiliation{Center for Theoretical Physics, Department of Physics and Astronomy, Seoul National University, Seoul 08826, Korea}

\author{Alessandro Lupi}
\affil{DiSAT, Universit\`a degli Studi dell'Insubria, via Valleggio 11, I-22100 Como, Italy}
\affil{Dipartimento di Fisica ``G. Occhialini'', Universit\`a degli Studi di Milano-Bicocca, I-20126 Milano, Italy}

\author[0000-0001-7457-8487]{Kentaro Nagamine}
\affiliation{Department of Earth and Space Science, Graduate School of Science, Osaka University, Toyonaka, Osaka, 560-0043, Japan}
\affiliation{Kavli IPMU (WPI), University of Tokyo, 5-1-5 Kashiwanoha, Kashiwa, Chiba, 277-8583, Japan}
\affiliation{Department of Physics \& Astronomy, University of Nevada Las Vegas, Las Vegas, NV 89154, USA}

\author[0000-0002-3764-2395]{Johnny W. Powell}
\affil{Department of Physics, Reed College, Portland, OR 97202, USA}

\author[0000-0001-5510-2803]{Thomas R. Quinn}
\affil{Department of Astronomy, University of Washington, Seattle, WA 98195, USA}

\author{Yves Revaz}
\affil{Institute of Physics, Laboratoire d'Astrophysique, \'{E}cole Polytechnique F\'{e}d\'{e}rale de Lausanne (EPFL), CH-1015 Lausanne, Switzerland}

\author{Ikkoh Shimizu}
\affil{Shikoku Gakuin University, 3-2-1 Bunkyocho, Zentsuji, Kagawa, 765-8505, Japan}

\author{H\'{e}ctor Vel\'{a}zquez}
\affil{Instituto de Astronom\'{i}a, Universidad Nacional Aut\'{o}noma de M\'{e}xico, A.P. 70-264, 04510, Mexico, D.F., Mexico}

\author{the {\it AGORA} Collaboration}
\affiliation{\rm \url{http://www.AGORAsimulations.org}}


\begin{abstract}
The {\it AGORA} {\tt Cosmorun} \citep{RocaFabrega2021} is a set of hydrodynamical cosmological ``zoom-in'' simulations carried out within the {\it AGORA} High-resolution Galaxy Simulations Comparison Project \citep{2014ApJS..210...14K_short,2016ApJ...833..202K_short}. These simulations show the formation and evolution of a Milky Way-sized galaxy using eight of the most widely used numerical codes in the community ({\sc Art-I}, {\sc Enzo}, {\sc Ramses}, {\sc Changa}, {\sc Gadget-3}, {\sc Gear}, {\sc Gizmo} and {\sc Arepo}.). In this short report, we describe the public release of the raw output data from all of these simulations at $z = 8, 7, 6, 5, 4, 3, 2$ (plus at $z=1, 0$ when available), and several metadata files containing the halo centers, virial quantities, and merger trees. The data from even thinner timesteps will be released as soon as the upcoming collaboration papers (VII-IX) are submitted and accepted.
\end{abstract}



\vspace{10mm}

\section{The {\it AGORA} Initiative:  Past, Present, and Future} 

Since its launch in 2012, the {\it AGORA} High-resolution Galaxy Simulations Comparison Project ({\it Assembling Galaxies of Resolved Anatomy}) has taken aim at carefully comparing high-resolution galaxy simulations on multiple code platforms widely used in the contemporary galaxy formation research.\footnote{See the Project website at \url{http://www.AGORAsimulations.org/} for more information about the {\it AGORA} Collaboration. \label{agora-website}}  
The main goal of this initiative has been to ensure that physical assumptions are responsible for any success in galaxy formation simulations, rather than manifestations of particular numerical implementations, and by doing so, to collectively raise the predictive power of numerical galaxy formation studies. 
Over 160 individuals from over 60 different academic institutions worldwide are participating or participated in the collaborative effort of the Project.

The first result by the {\it AGORA} Collaboration \citep[][hereafter Paper I]{2014ApJS..210...14K_short} was our flagship paper which explained the philosophy behind the initiative and detailed the publicly available Project infrastructure we have assembled. 
Also described was the proof-of-concept test, in which we field-tested our infrastructure with a dark matter-only cosmological zoom-in simulation using the common initial condition \citep[generated with {\sc Music};][]{MUSIC}, finding a robust convergence amongst participating codes. 
In the second paper of the Project \citep[][hereafter Paper II]{2016ApJ...833..202K_short}, we focused on the evolution of an isolated Milky Way-mass galaxy. 
All participating codes shared the common initial condition \citep[generated with {\sc Makedisk};][]{gadget2}, common physics models \citep[e.g., radiative cooling and extragalactic ultraviolet background provided by the standardized package {\sc Grackle};][]{2017MNRAS.466.2217S}, and common analysis platform \citep[{\tt yt} toolkit;][]{yt}. 
Subgrid physics models such as Jeans pressure floor, star formation, supernova feedback energy, and metal production were carefully constrained across code platforms.
With a spatial resolution of 80 pc that resolves the scale height of the disk, we find that any intrinsic inter-code difference is small compared to the variations in input physics such as supernovae feedback. 
In the third paper \citep[][hereafter Paper III]{RocaFabrega2021} we presented the {\tt Cosmorun} project and the calibration process we designed to reduce the number of variables to account for when comparing results from the different code groups. For these simulations, we adopted most of the subgrid physics and simulation strategies developed for Paper II but including the most recent version of the {\sc Grackle} library, leaving only to the decision of each code group the stellar feedback strategy to be used. 
Following Paper III publication, the {\tt Cosmorun} data has been used to study many aspects of the formation and evolution of MW-size galaxies focusing on the differences between the participant codes and their stellar feedback strategies. In particular, we presented the analysis of changes in the merger history, the stellar disk properties and the stellar metallicity distribution in \citet[][hereafter Paper IV]{RocaFabrega2024},  the satellites number counts in \citet[][hereafter Paper V]{Jung2024} and the circumgalactic medium (CGM) in \citet[][hereafter Paper VI]{Strawn2024}. In Table~\ref{tab:1} we include a summary of the {\it AGORA} papers published until August 2024. Currently, many new sub-projects are using the {\tt Cosmorun} to study the properties and formation of stellar disks, the quenching of satellites, and major merger properties, among many others.
After twelve years of workshops and teleconferences, and with common infrastructures built together, the {\it AGORA} Collaboration has become a successful open forum where users of different simulation codes can talk to and learn from one another, promoting collaborative and reproducible research essential in any scientific community.    
\begin{table*}
\vspace*{1mm}
\footnotesize
\caption{Summary of the published {\it AGORA} papers until August 2024.}
\vspace*{-2mm}
\centering
\begin{tabular}{c || c | c | c | c | c | c }
\hline\hline
\citet{2014ApJS..210...14K_short} & Paper I & Zoom-in Cosmo. & L$_{Box}$ = 60h$^{-1}$cMpc&  M$_{200c}$=1.7$\times$10$^{11}$M$\odot$ & Dark Matter Only & 9 codes \\ 
\hline
\citet{2016ApJ...833..202K_short} & Paper II & Isolated box& L$_{Box}$ = 1.074 Mpc &  M$_{200c}$=1.074$\times$10$^{12}$M$\odot$& Common stellar feedback & 9 codes\\
\hline
\citet{RocaFabrega2021} & Paper III & Zoom-in Cosmo. & L$_{Box}$ = 60h$^{-1}$cMpc&  M$_{200c}$=1.0$\times$10$^{12}$M$\odot$ & Calibrated feedback & 7-8 codes\\
\citet{RocaFabrega2024} & Paper IV & \multicolumn{5}{c}{Study of the merger history and disk morphology down to z=2 and below in the Paper III simulation ({\tt Cosmorun})}\\
\citet{Jung2024} & Paper V &  \multicolumn{5}{c}{Analysis of the satellites population in the {\tt Cosmorun} simulation}\\
\citet{Strawn2024} & Paper IV & \multicolumn{5}{c}{Analysis of the Circumgalactic Medium in the {\tt Cosmorun} simulation} \\
\hline
\end{tabular}
\label{tab:1}
\vspace*{1mm}
\end{table*}

\section{Public Release of {\tt Cosmorun} data} 

All code groups started their simulations from a common initial condition (IC) generated with {\sc Music} \citep[][]{Hahn2011},\footnote{The website is \url{https://www-n.oca.eu/ohahn/MUSIC/}.} and run by keeping physics prescriptions the same for all codes (e.g., gas cooling and heating, star formation parameters), although some variations were made in each code \citep[see][for details on these differences]{RocaFabrega2021}. Only the decision concerning the stellar feedback prescription and metal production to be used was left to each code group. 
Code groups were asked to use a prescription close to the most widely-used practice in each code community. 
Spatial resolution was $\sim$80 physical pc at $z = 4$ to resolve the internal structure of a target halo, and to make our physics prescriptions less reliant on platform-specific models.

Here we provide the simulation snapshots used in the analysis of Paper III plus the one at $z = 3, 2$ and $1, 0$ when available (see Appendix~\ref{appendix} for a detailed description of the released data). We also provide the metadata of the centers, several virial quantities, and the results of applying the {\tt rockstar} halo finder to each one of the datasets.
The cohort of widely-used, state-of-the-art galaxy simulation codes who contributed to this release includes:  the Lagrangian smoothed particle hydrodynamics codes {\sc Changa} \citep[e.g.,][]{Menon15}, {\sc Gadget-3} \citep[e.g.,][]{2012MNRAS.419.1280C, 2017MNRAS.466..105A,Shimizu2019,Nagamine2021}, and {\sc Gear} \citep[e.g.,][]{revaz_computational_2016}, the Eulerian adaptive mesh refinement codes {\sc Art-I} \citep[e.g.,][]{2014MNRAS.442.1545C}, {\sc Enzo} \citep[e.g.,][]{2014ApJS..211...19B} and {\sc Ramses} \citep[e.g.,][]{ramses}, the moving-mesh code {\sc Arepo} \citep[e.g.,][]{2020ApJS..248...32W} and the mesh-free finite-volume Godunov code {\sc Gizmo} \citep[e.g.,][]{hopkins2015}. 

This release is intended to encourage the community to re-run the {\tt Cosmorun} ICs both after applying the provided calibration procedure or not and using their favourite stellar feedback. Our goal is to enlarge the library of {\tt Cosmorun} models thus enhancing our insight on how variations in the baryonic physics processes affect the formation of MW size galaxies. The coordination of the Collaboration and its members are at their disposal to help with the calibration process and with running new models.

The simulation data and several of the common analysis scripts in {\tt yt} are available through the Project website\footnote{\url{http://www.AGORAsimulations.org/} or \url{http://sites.google.com/site/santacruzcomparisonproject/blogs/quicklinks/}}. Additionally, the authors of the published {\it AGORA} papers (see Table~\ref{tab:1}) will be happy to provide help to users interested in analyzing the released data.
Also available in the same link are isolated and cosmological initial conditions generated by the {\it AGORA} Collaboration for galaxy simulations, and the links to the key software.  
We encourage numerical galaxy formation community members to use these resources freely for their research.

\bibliography{sample63}

\appendix
\section{Details of the released data}\label{appendix}
In Table~\ref{tab:1_app} we provide the information on the snapshot number at each redshift for each one of the participant codes. We also provide the link to the data for each code, below the table.\\

In the metadata folder\footnote{\href{https://users.flatironinstitute.org/~chayward/agora\_public\_release/PaperIV/Metadata/}{https://users.flatironinstitute.org/$\sim$chayward/agora\_public\_release/PaperIV/Metadata/}} you will find three different kind of files and a subfolder (\href{https://users.flatironinstitute.org/~chayward/agora_public_release/PaperIV/Metadata/MergerTrees_rockstar/}{\sc MergerTrees\_rockstar}) which contains the merger trees generated using {\tt rockstar} \citep{Behroozi2013} and {\tt consistent-trees} \citep{ConsistentTrees2012}. All filenames contain the name of the corresponding code except for {\sc Changa} which is labeled as  {\tt CHANGA\_6.0e51erg\_NSB} for the {\sc Changa-T} and {\tt CHANGA\_3.5e51erg\_SB} for the fiducial {\sc Changa}. Here we describe the content of each file and of the merger-trees folder:\\
\begin{itemize}
\item {\tt centers\_CODE\_final\_noID.txt}: This file contains the centers of the main halo inside the simulated cosmological box in physical kiloparsecs. The data is structured as follows: redshift, snapnum (see Table~\ref{tab:1_app}), [x,y,z].\\
\item {\tt Rvir\_cent\_CODE.txt}: This file contains similar information to the previous one but in a simpler format, including information on the R$_{200c}$. The data is structured as follows: redshift, snapnum (see Table~\ref{tab:1_app}), x, y, z, R$_{200c}$ (in physical kpc).\\
\item {\tt Rvir\_cent\_others\_CODE.txt}: This file contains information about many general properties of the central halo. The data is structured as follows: redshift, snapnum (see Table~\ref{tab:1_app}), R$_{200c}$ [kpc], M$_{200c}$ [M$_{\odot}$], M$_{*}$(R$<$R$_{200c}$) [M$_{\odot}$], M$_{*,gal}$(R$<0.15$R$_{200c}$) [M$_{\odot}$], R$_{1/2}$ [kpc], M$_{tot}$(R$<$R$_{1/2}$) [M$_{\odot}$], Z$_{*}$ [Z$_{\odot}$], R$_{1/2*,gal}$ [kpc], M$_{*,gal}$(R$<$R$_{1/2*}$) [M$_{\odot}$], Z$_{*,gal}$ [Z$_{\odot}$].\\
\item {\sc MergerTrees\_rockstar}: In this folder you will find a sub-folder containing the merger for each one of the participant codes except for the fiducial {\sc Changa}. The files can be read and analysed using the  {\tt ytree}\footnote{\href{https://ytree.readthedocs.io/en/latest/}{https://ytree.readthedocs.io/en/latest/}} tool provided by the {\tt yt}\footnote{\href{https://yt-project.org}{https://yt-project.org}} community.  
\end{itemize}

\begin{table*}
\begin{center}
\caption{In this table we provide information on the snapshot number at each redshift and for each one of the participating codes. We also provide the links to the data, below.}
\label{tab:1_app}
\begin{tabular}{c || c | c | c | c | c | c | c | c | c | c | c | c | c | c | c | c }
\hline\hline
CODE & \multicolumn{16}{c}{Snapshot number as in the metadata files for z=15 to 0} \\
 & 15 & 14 & 13 & 12 & 11 & 10 & 9 & 8 & 7 & 6 & 5 & 4 & 3 & 2 & 1 & 0 \\
 \hline
{\sc ART-I}$^{i}$ & 1729 & 1765 & 1802 & 1841 & 1882 & 1924 & 1968 & 2014 & 2071 & 2153 & 2245 & 2385 & 2580 & 2833 & 3273 & 4078 \\ 
{\sc ENZO}$^{ii}$ & 000 & 009 & 019 & 029 & 039 & 051 & 063 & 077 & 092 & 109 & 129 & 152 & 180 & 213 & 247 & 347 \\ 
{\sc RAMSES}$^{iii}$ & 002 & 011 & 020 & 029 & 041 & 053 & 065 & 079 & 094 & 111 & 131 & 154 & 182 & 220 & - & -\\
{\sc CHANGA}$^{iv}$ & 0150 & 0167 & 0186 & 0208 & 0238 & 0271 & 0314 & 0370 & 0442 & 0542 & 0688 & 0904 & 1264 & 1932 & 3440 & -\\
{\sc CHANGA-T}$^{v}$ & 0142 & 0159 & 0177 & 0199 & 0224 & 0258 & 0298 & 0352 & 0422 & 0516 & 0654 & 0860 & 1202 & 1838 &-& -\\
{\sc GADGET-3}$^{vi}$ & 001 & 009 & 017 & 026 & 036 & 046 & 057 & 070 & 084 & 100 & 119 & 141 & 168 & 203 & 252 & -\\
{\sc GEAR}$^{vii}$ & 001 & 022 & 046 & 071 & 098 & 128 & 161 & 196 & 236 & 281 & 333 & 394 & 469 & 566 & 701 & -\\ 
{\sc AREPO-T}$^{viii}$ & 000 & 009 & 017 & 026 & 036 & 046 & 057 & 070 & 085 & 100 & 119 & 142 & 169 & 203 & 152 & 336 \\
{\sc GIZMO}$^{ix}$ & 000 & 009 & 019 & 029 & 039 & 051 & 063 & 077 & 092 & 109 & 129 & 152 & 180 & 215 &- & -\\ 
\hline
\end{tabular}
\end{center}
\tablenotetext{i}{\href{https://users.flatironinstitute.org/~chayward/agora\_public\_release/PaperIV/ART-I/}{\ \ \  https://users.flatironinstitute.org/$\sim$chayward/agora\_public\_release/PaperIV/ART-I/}}
\tablenotetext{ii}{\href{https://users.flatironinstitute.org/~chayward/agora\_public\_release/PaperIV/ENZO/}{\ \ \    https://users.flatironinstitute.org/$\sim$chayward/agora\_public\_release/PaperIV/ENZO/}}
\tablenotetext{iii}{\href{https://users.flatironinstitute.org/~chayward/agora\_public\_release/PaperIV/RAMSES/}{\ \ \     https://users.flatironinstitute.org/$\sim$chayward/agora\_public\_release/PaperIV/RAMSES/}}
\tablenotetext{iv}{\href{https://users.flatironinstitute.org/~chayward/agora\_public\_release/PaperIV/CHANGA/SB/output/}{\ \ \ https://users.flatironinstitute.org/$\sim$chayward/agora\_public\_release/PaperIV/CHANGA/SB/output/}}
\tablenotetext{v}{\href{https://users.flatironinstitute.org/~chayward/agora\_public\_release/PaperIV/CHANGA/nSB/output/}{\ \ \ https://users.flatironinstitute.org/$\sim$chayward/agora\_public\_release/PaperIV/CHANGA/nSB/output/}}
\tablenotetext{vi}{\href{https://users.flatironinstitute.org/~chayward/agora\_public\_release/PaperIV/GADGET3/}{\ \ \ https://users.flatironinstitute.org/$\sim$chayward/agora\_public\_release/PaperIV/GADGET3/}}
\tablenotetext{vii}{\href{https://users.flatironinstitute.org/~chayward/agora\_public\_release/PaperIV/GEAR/}{\ \ \ https://users.flatironinstitute.org/$\sim$chayward/agora\_public\_release/PaperIV/GEAR/}}
\tablenotetext{viii}{\href{https://users.flatironinstitute.org/~chayward/agora\_public\_release/PaperIV/AREPO/}{\ \ \ https://users.flatironinstitute.org/$\sim$chayward/agora\_public\_release/PaperIV/AREPO/}}
\tablenotetext{ix}{\href{https://users.flatironinstitute.org/~chayward/agora\_public\_release/PaperIV/GIZMO/}{\ \ \ https://users.flatironinstitute.org/$\sim$chayward/agora\_public\_release/PaperIV/GIZMO/}}
\end{table*}



\end{document}